\documentclass{aastex631}
\usepackage{amsmath} 
\usepackage{amsbsy}

\begin{document}

\title{Radial Evolution of a Corotating Interaction Region from a High-Resolution MHD Simulation}

\correspondingauthor{Liping Yang, Fang Shen}
\email{lpyang@swl.ac.cn, fshen@spaceweather.ac.cn}

\author{Mengxuan Ma}
\affiliation{SIGMA Weather Group, State Key Laboratory of Solar Activity and Space Weather, National Space Science Center, Chinese Academy of Sciences, Beijing 100190, People's Republic of China}
\affiliation{College of Earth and Planetary Sciences, University of Chinese Academy of Sciences, Beijing 100049, People's Republic of China}

\author{Liping Yang}
\affiliation{School of Geophysics and Information Technology, China University of Geosciences (Beijing), China}
 
\author{Fang Shen}
\affiliation{SIGMA Weather Group, State Key Laboratory of Solar Activity and Space Weather, National Space Science Center, Chinese Academy of Sciences, Beijing 100190, People's Republic of China}
\affiliation{College of Earth and Planetary Sciences, University of Chinese Academy of Sciences, Beijing 100049, People's Republic of China}

\author{Xueshang Feng} 
\affiliation{SIGMA Weather Group, State Key Laboratory of Solar Activity and Space Weather, National Space Science Center, Chinese Academy of Sciences, Beijing 100190, People's Republic of China} 

\author{Xinyi Ma}
\affiliation{SIGMA Weather Group, State Key Laboratory of Solar Activity and Space Weather, National Space Science Center, Chinese Academy of Sciences, Beijing 100190, People's Republic of China} 
\affiliation{College of Earth and Planetary Sciences, University of Chinese Academy of Sciences, Beijing 100049, People's Republic of China}

\author{Wangning Zhang}
\affiliation{SIGMA Weather Group, State Key Laboratory of Solar Activity and Space Weather, National Space Science Center, Chinese Academy of Sciences, Beijing 100190, People's Republic of China}
\affiliation{School of Geophysics and Information Technology, China University of Geosciences (Beijing), People’s Republic of China}

\author{Xiao Zhang}
\affiliation{SIGMA Weather Group, State Key Laboratory of Solar Activity and Space Weather, National Space Science Center, Chinese Academy of Sciences, Beijing 100190, People's Republic of China}
\affiliation{College of Earth and Planetary Sciences, University of Chinese Academy of Sciences, Beijing 100049, People's Republic of China}

\accepted{by ApJ}

\begin{abstract}  
Corotating interaction regions (CIRs) are large-scale structures formed by fast--slow solar wind interaction, playing a crucial role in heliospheric dynamics and space weather. However, the evolution of individual CIRs and their full three-dimensional morphology remain poorly understood. We investigate the radial evolution of a single CIR using a high-resolution simulation combined with a backmapping method. The simulation reproduces the major plasma and magnetic signatures observed by Wind at 1~AU and by MAVEN and Tianwen-1 at $\sim$1.6~AU. The CIR exhibits a warped, shell-like spiral structure tilted relative to the heliospheric current sheet, consistent with theoretical predictions. During its developing stage, geometric changes are modest, and the radial scalings of velocity ($V_r \propto r^{0.02}$), density ($N \propto r^{-2.00}$), magnetic field ($B_{\rm total} \propto r^{-1.74}$), and temperature ($T \propto r^{-0.97}$) are comparable to those of the background solar wind. In the mature stage, the CIR drifts in longitude and expands toward high latitudes, while its longitudinal and latitudinal widths decrease with distance. The full-domain scalings change markedly to $V_r \propto r^{0.13}$, $N \propto r^{-1.91}$, $B_{\rm total} \propto r^{0.35}$, and $T \propto r^{-0.52}$, differing from both the background wind and the ecliptic-plane CIR. This study quantifies the radial evolution of a CIR from formation to maturity and highlights the importance of its three-dimensional structure in characterizing its morphology and dynamics.
\end{abstract}



\section{Introduction}

The solar wind, a continuous outflow of magnetized plasma from the Sun, fills the heliosphere and mediates the transport of solar activity into interplanetary space \citep{Parker1958}. Embedded within this flow are transient coronal mass ejections (CMEs) and recurrent corotating interaction regions (CIRs), which constitute the two dominant drivers of large-scale heliospheric variability and space weather disturbances \citep{Gopalswamy2009, Yuan2013, Zhang2021, Dai2023}. CMEs are impulsive eruptions releasing energies of $10^{20}$--$10^{25}$~J \citep{Chen_2011}, whereas CIRs are quasi-stationary structures formed by the interaction between fast and slow solar wind streams and can persist over multiple solar rotations \citep{Richardson2018}.

CIRs arise when high-speed solar wind streams originating from coronal holes overtake preceding slow solar wind, producing large-scale compression regions \citep{Gosling1999, Richardson2018}. These interaction regions are typically bounded by forward and reverse pressure waves, which can steepen into shocks at larger heliocentric distances \citep{Belcher1971, Jian2006}. The resulting compression enhances plasma density, magnetic field strength, and temperature, making CIRs efficient sites for particle acceleration and important contributors to recurrent geomagnetic activity \citep{Giacalone1997, Borovsky2006, Dai2024}.

The radial evolution of CIR properties has been the subject of extensive observational studies. Observations from the Helios 1 and 2 spacecraft revealed that, alongside velocity steepening, the density, temperature, pressure, and magnetic field at the CIR interface all increase with radial distance between 0.3-0.4 AU and 0.9-1.0 AU \citep{Richter1986, Forsyth1999}. Subsequent comparisons of CIR observations from Pioneer, Venus Orbiter, and the Advanced Composition Explorer (ACE)/Wind show that the occurrence rate of CIRs with sharp stream interfaces increased from 9\% at 0.7 AU to 21\% at 1 AU, while the velocity variation exhibited little change over this distance range \citep{Jian2008}. A fortuitous conjunction between Parker Solar Probe (PSP) and the Solar Terrestrial Relations Observatory-Ahead (STEREO-A) enabled a detailed assessment of the radial evolution of a single CIR, whose bulk plasma and magnetic field signatures evolve in a manner consistent with these prior trends \citep{Allen2021a, Allen2021b}.

More recently, the study of CIR radial evolution has been extended to 1.6 AU using data from the Mars Atmosphere and Volatile Evolution (MAVEN) and Tianwen-1 spacecraft \citep{Geyer2021, Zhong2024, Henderson2025}. These observations indicate that at Mars, the occurrence rate of fast forward shocks is three times higher, and the high-speed streams are broader, while the magnetic field and total pressure are evidently enhanced compared to those at Earth. Furthermore, comparative analyses of CIR signatures observed by STEREO-A/B and ACE/Wind suggest that the probability of repeatedly detecting a given CIR is not significantly influenced by the latitudinal separation between the observing spacecraft \citep{Allen2020, Chi2022}.

Despite these advances in characterizing the average CIR profile as a function of radial distance, our understanding of individual CIR evolution remains limited. This is primarily due to observational constraints: only a handful of events have been tracked by spacecraft at different radial distances while sharing the same heliographic longitude and latitude \citep{Milo2025}, and most studies rely on single-point measurements that offer merely one-dimensional cuts through the inherently three-dimensional CIR structure \citep{Wimmer1997}. As a result, it remains unclear whether the empirical scaling laws for the radial evolution of CIR parameters, derived from multi-spacecraft composites, are applicable to any given single event. Furthermore, the global morphology of CIRs, including their latitudinal extent and longitudinal width, remains poorly constrained. These uncertainties pose a critical challenge for reliably predicting CIR properties from one heliocentric location to another.

Magnetohydrodynamic (MHD) simulations offer a complementary approach that directly addresses these observational limitations. By providing continuous three-dimensional coverage of the simulation domain, they enable both the tracking of individual CIRs across arbitrary radial distances and the characterization of their full global morphology \citep{Feng2020}. Previous simulation studies have focused on successfully reproducing key large-scale features of CIRs, including the increase in wind speed, the increase of total pressure with a gradual decrease on both flanks of the compression region, the non-monotonic variation of proton density (rising then falling), the elevation of proton temperature, and the compression of magnetic field strength \citep{Pizzo1982, Pizzo1994, Odstrcil2003, Riley2007, Jian2011, Riley2012, Pomoell2018, Wexler2023, Aguilar2024, Henderson2025b}. However, despite these successful validations, a systematic investigation of how the physical and geometric properties of individual CIRs evolve with radial distance remains notably absent from the literature.

Three recent simulation studies are particularly close to the present work and therefore deserve further discussion. \citet{Wexler2023} studied an MHD heliospheric simulation containing a well-defined stream interaction region (SIR) and compared the simulated density and magnetic field enhancements, as well as the accompanying magnetic polarity reversal, with Wind measurements at 1~AU; the good agreement obtained demonstrates that the global MHD solution is able to reproduce the in situ signature of the interaction region. \citet{Wraback2024} performed a high-resolution stream-aligned MHD simulation of the compressive interaction region encountered during Parker Solar Probe's first perihelion and compared the simulated plasma and magnetic field parameters simultaneously with in situ measurements from Parker Solar Probe, STEREO-A, and near-Earth spacecraft; by additionally treating the energy partitioning between the parallel and perpendicular ions, they were able to predict the increase of the wave energy at the compression region as well as the proton temperature anisotropy. \citet{Zhang2024} simulated the solar wind near Mars over the same interval with a three-dimensional total variation diminishing MHD model and assimilated the MAVEN and Tianwen-1 in situ measurements through a Kalman filter, an approach that aims at an accurate local reconstruction of the near-Mars solar wind rather than at the three-dimensional structure and radial evolution of an individual CIR considered here. These studies demonstrate the ability of global MHD models to reproduce CIR signatures at the heliocentric distances sampled by the available spacecraft; a systematic description of the three-dimensional geometry of one and the same CIR and of its radial evolution from the inner heliosphere to beyond 1.6~AU has, however, been lacking.

In this work, we aim to partially address this gap by simulating a CIR event observed by Wind at 1 AU and by MAVEN and Tianwen-1 at about 1.6 AU, using the three-dimensional (3D) Solar-Interplanetary (SIP) adaptive mesh refinement (AMR) space-time conservation element and solution element (CESE) solar wind model
(SIP-AMR-CESE Solar Wind Model) \citep{Feng2010, Feng2012, Yang2012}. To reproduce the low-latitude high-speed solar wind essential for CIR formation, the simulation applies enhanced solar wind heating along open field lines originating from low-latitude solar wind sources, identified by a two-step backmapping technique. The AMR technique is also employed to capture the CIR with high resolution, enabling the quantification of the radial evolution of its geometric properties, including longitudinal and latitudinal extents. In addition, we evaluate the radial evolution of key parameters within the CIR, such as speed, density, magnetic field strength, and temperature, to derive empirical scaling laws. The remainder of this paper is organized as follows: Section 2 describes the numerical model and boundary conditions, Section 3 presents the simulation results and their comparison with multi-spacecraft observations, and Section 4 summarizes the main findings and discusses their implications.

\section{Model and Setting}

\subsection{SIP-AMR-CESE Solar Wind Model}

In this study, we employ the three-dimensional SIP-AMR-CESE solar wind model to simulate the ambient solar wind and the radial evolution of CIRs from the low corona to the heliosphere \citep{Feng2012, Yang2012, Yang2023, Ma2024}. The model solves the standard set of ideal MHD equations in conservative form, governing the conservation of mass, momentum, magnetic flux, and total energy. The space-time conservation element and solution element (CESE) scheme used here is shock-capturing, so that discontinuities such as shocks can be represented when they form. To account for the heating and acceleration of the solar wind, the effects of magnetic field topology, specifically the expansion factor and the minimum angular
separation between an open field foot point and its nearest coronal hole boundary, are incorporated into the volumetric heating term $Q$ \citep{Feng2010, Yang2021, Zhang2026}.

The computational domain is discretized using the six-component grid system proposed by \citet{Feng2010}, which divides the spherical shell into six mutually matched blocks, effectively avoiding the pole singularities and grid convergence issues inherent to conventional spherical coordinates. The domain spans  $1~Rs \leq r\leq 500~Rs$, $0^\circ \leq \theta \leq 180^\circ$, and $0^\circ \leq \phi \leq 360^\circ$. Here, $Rs$ is solar radius, $\theta$ is  colatitude angle, and $\phi$ is  longitude angle.  

The ambient solar wind is initialized using the Parker hydrodynamic solar wind solution, which prescribe the initial plasma density $\rho$, gas pressure $p$, and plasma velocity $\mathbf{u}$. 
The temperature and number density at the solar surface are set to  $1.45\times10^6$ K and $2\times10^8$ cm$^{-3}$, respectively.
The initial magnetic field is specified as a potential field derived from the Global Oscillation Network Group (GONG) synoptic magnetogram for Carrington Rotation 2251.

To accurately resolve the multiscale structures associated with the CIR while maintaining computational efficiency, the adaptive mesh refinement (AMR) technique is employed throughout the simulation. The refinement procedure is guided by both physical quantities and the spatial location of the target CIR. Specifically, regions exhibiting enhanced total pressure, including both thermal pressure and magnetic pressure, are identified as potential refinement regions because they correspond to the compression interface between the fast and slow solar wind streams. In addition, the spatial position of the target CIR obtained from the previous time step is used to continuously track the evolution of the interaction region. Based on these refinement criteria, the computational mesh is dynamically refined up to five AMR levels around the target CIR.  Each refinement doubles the cell resolution, so that inside the CIR the radial cell width is approximately 1.3, 1.5 and 1.8~Rs at 0.67, 1.0 and 1.6~AU, respectively. This strategy enables the compression region, stream interface, and associated plasma gradients to be resolved with significantly higher spatial resolution, while avoiding unnecessary refinement in the remaining computational domain and thereby substantially improving computational efficiency.

\subsection{Solar Wind Source Mapping and Heating Setting}

\begin{figure*}[ht!]
\centering
\includegraphics[width=\textwidth]{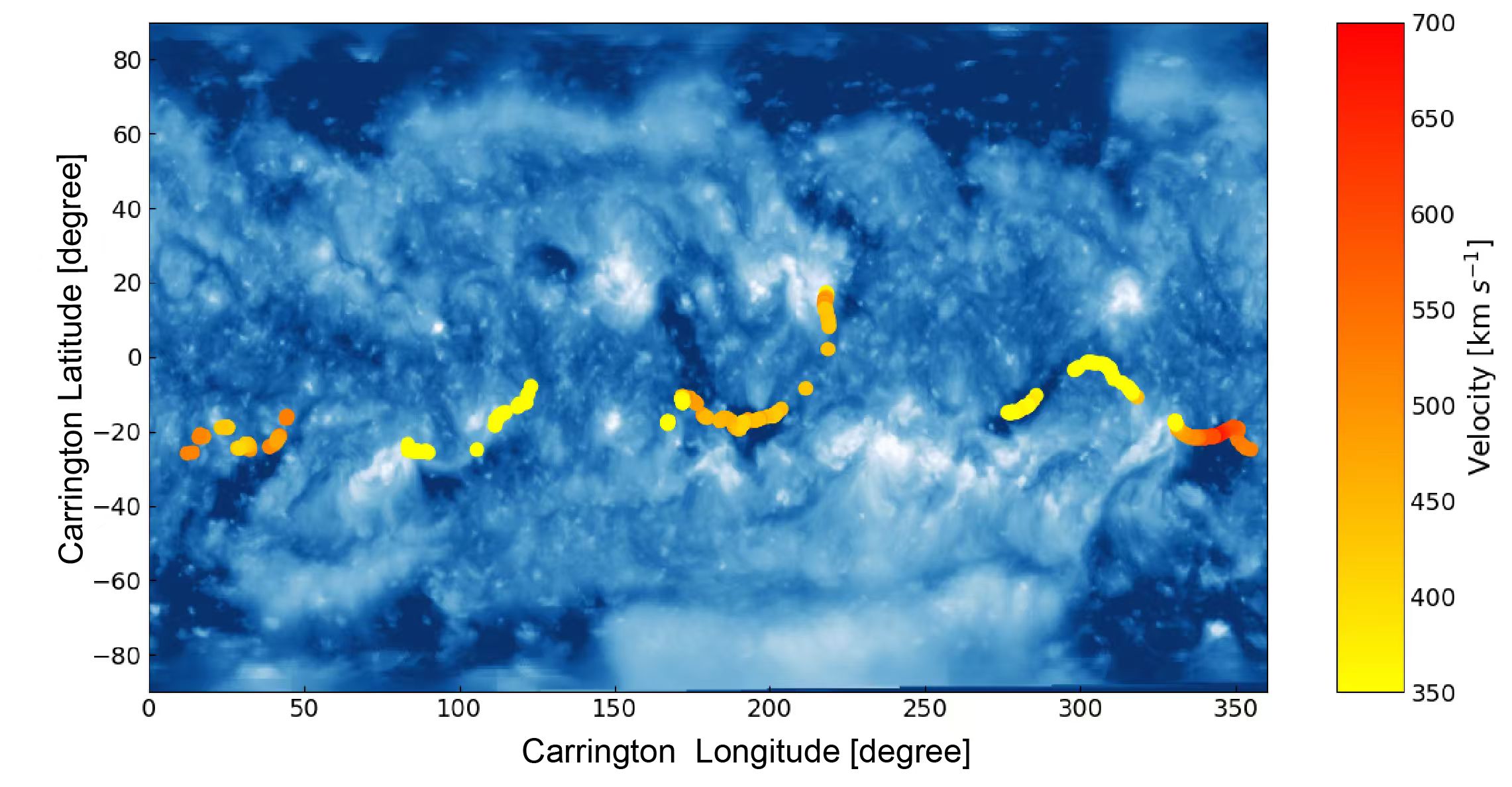}
\caption{STEREO-A/SECCHI EUVI 195~\AA{} synoptic map showing the source regions of the solar wind measured by WIND during CR2251, 
with each source region marked by a colored dot indicating the solar wind speed.}
\label{fig:pfss_structure}
\end{figure*}

To reproduce the high-speed solar wind measured by WIND, we first map solar wind parcels at 1~AU back to their source regions on the Sun using the two-step backmapping method \citep{Hou2024, Ma2025}. This approach establishes a direct linkage between in-situ plasma properties and their coronal origins, providing essential insight into the formation and subsequent evolution of CIRs from the Sun to interplanetary space. The backmapping results are shown in Figure~\ref{fig:pfss_structure}, which indicates that the majority of the solar wind, particularly the high-speed streams, originates from low-latitude small-scale coronal holes.

Based on the backmapping results, which identify both the origins and the propagation paths of the high-speed solar wind measured by WIND, we introduce an additional volumetric heating term 
$Q_2$ to augment the baseline heating term $Q$, thereby mimicking the heating effects associated with the fast wind. Specifically, we set a flag to 1 along the field lines that trace from the source regions of the high-speed streams to 1~AU, which activates the additional heating term $Q_2$. The heating rate is typically parameterized as a function of radial distance to ensure that the simulated solar wind reaches the observed speeds at 1~AU.

\section{Results}

\subsection{Validation of the Simulation Results}

\begin{figure*}[ht!]
\centering
\includegraphics[width=0.8\textwidth]{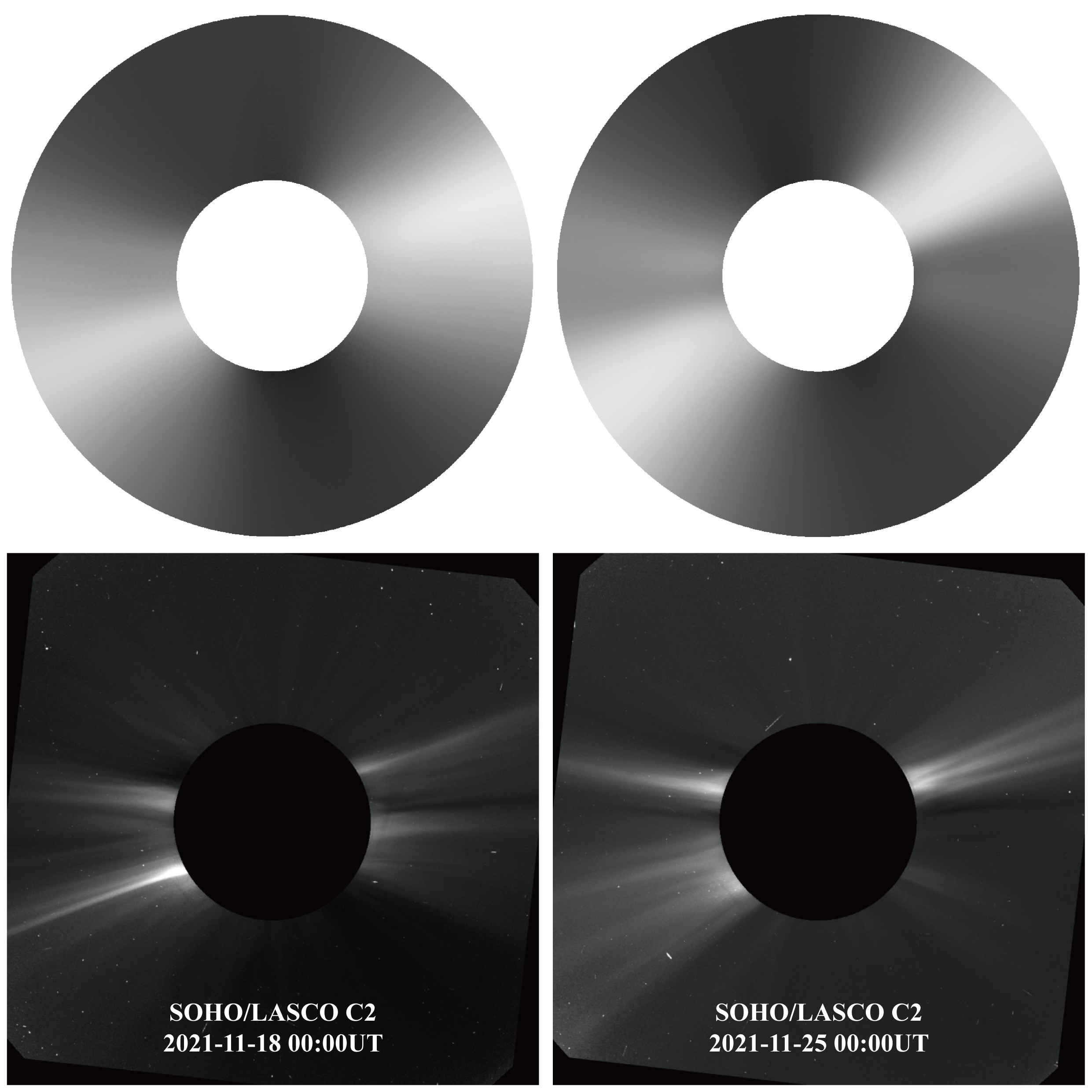}
\caption{
Comparison of the white-light coronagraph images between the simulation and the SOHO/LASCO C2 observations for CR2251.
The upper panels show the synthetic white-light images generated from the MHD simulation at Carrington longitudes of $0^\circ-180^\circ$ (left) and $90^\circ-270^\circ$ (right), while the lower panels present the corresponding SOHO/LASCO C2 observations obtained on 2021 November 18 at 04:00~UT and 2021 November 25 at 00:00~UT, respectively.
}
\label{fig:corona}
\end{figure*}

Figure~\ref{fig:corona} compares the synthesized white-light coronagraph images from our simulation with the SOHO/LASCO C2 observations. The simulated streamer belt and large-scale coronal structures show good agreement with the observed coronal morphology from different viewing longitudes, indicating that the simulated coronal magnetic field, together with the solar wind solution, provides a reliable representation of the global coronal environment. This reconstructed coronal solution thus serves as a realistic starting point for the subsequent heliospheric propagation of the solar wind.

\begin{figure*}[ht!]
\centering
\includegraphics[width=\textwidth]{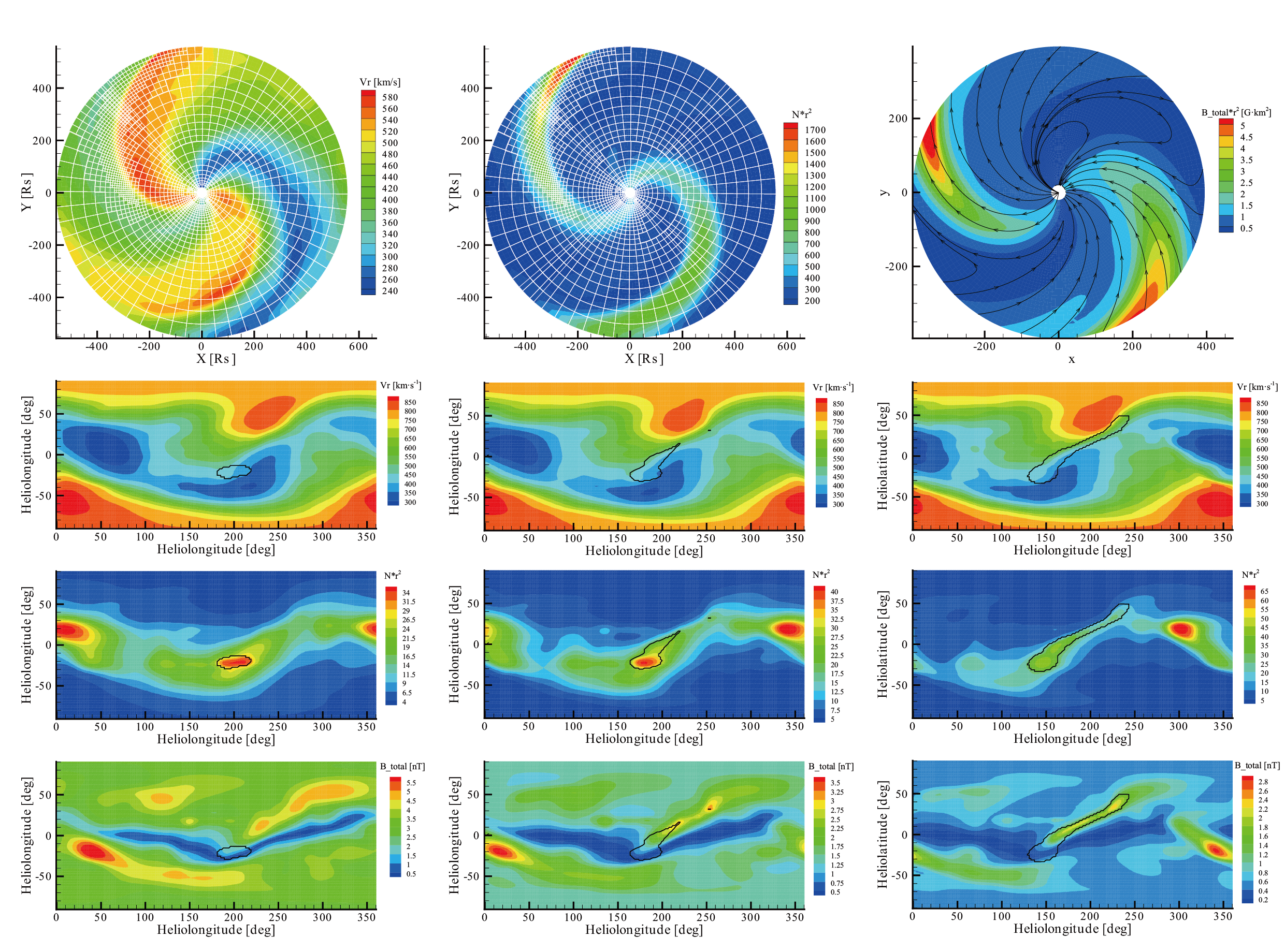}
\caption{
Simulated heliospheric solar wind during CR2251. The first row presents the distributions of radial velocity ($V_r$), normalized plasma density ($N*r^2$), and total magnetic field strength ($B_\textrm{total}$) in the ecliptic plane, where the white quadrilaterals denote the computation blocks, each subdivided into $6\times6\times6$ grid cells. The remaining rows show the corresponding distributions at heliocentric distances of 0.67, 1.0, and 1.6 AU, respectively, with black curves identifying the CIR. 
}
\label{fig:ecliptic}
\end{figure*}

Figure~\ref{fig:ecliptic} presents the simulated distributions of radial velocity, plasma density, and magnetic field strength in the ecliptic plane, together with radial slices at selected heliocentric distances. Two high-speed streams, which extend from the polar regions, are present in the ecliptic plane, and their interaction with the preceding slow solar wind naturally forms a compressed interaction region characterized by enhanced plasma density and magnetic field. As the solar wind propagates outward, these compression regions gradually develop into well-defined CIRs that exhibit the characteristic spiral morphology and grow increasingly prominent with distance. One of the two CIRs is selected for high-resolution analysis in the following sections.

\begin{figure*}[ht!]
\centering
\includegraphics[width=\textwidth]{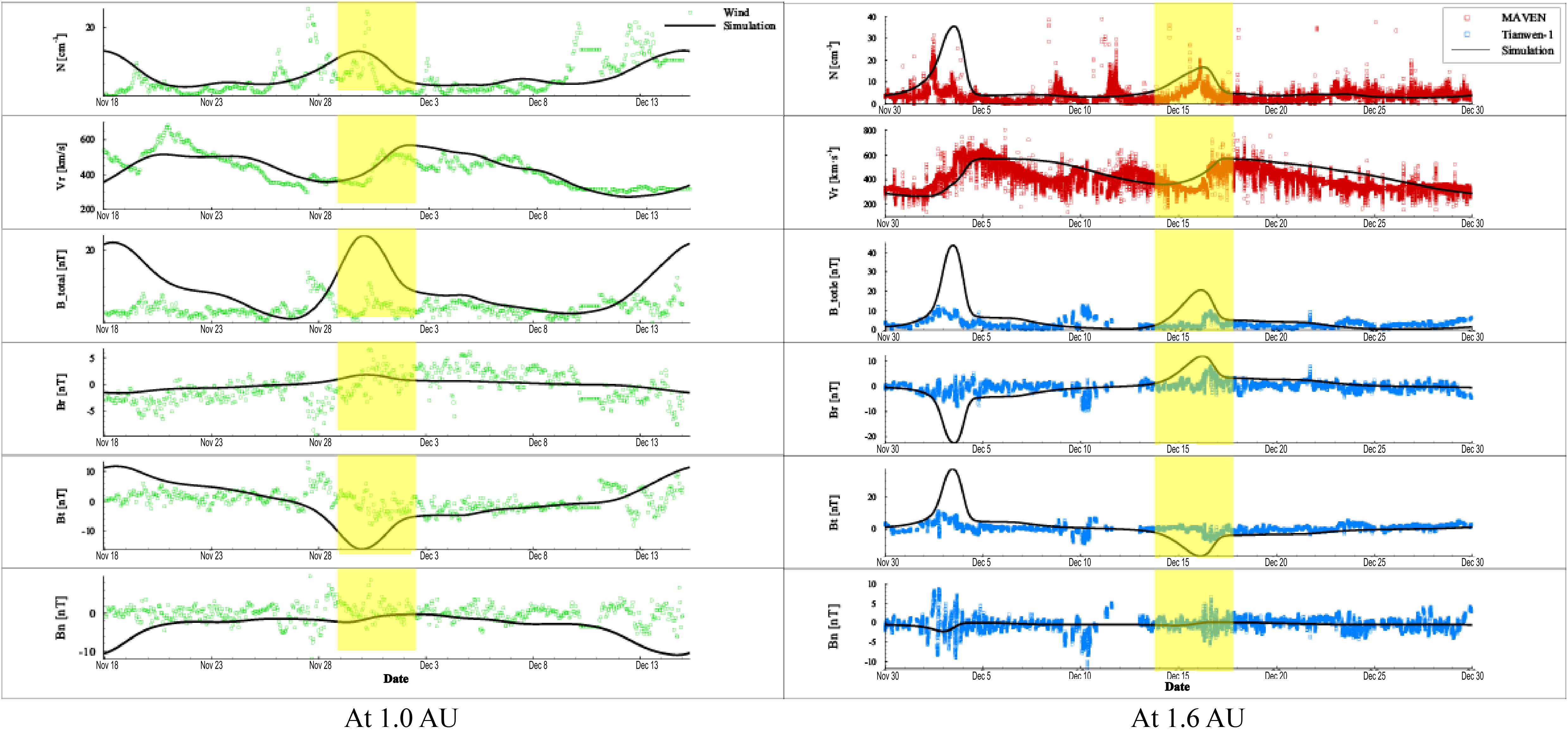}
\caption{
Comparison between the simulated solar wind parameters (black lines) and in-situ measurements from WIND (green lines, left panels), and from MAVEN (red lines, right panels) and Tianwen-1 (blue lines, right panels). From top to bottom, the panels show the plasma density, radial velocity, total magnetic field strength, and magnetic field components. The yellow shaded regions mark the intervals during which the target CIR resolved at high resolution encountered each spacecraft.  
}
\label{fig:validation}
\end{figure*}

The reliability of the simulation is further evaluated through comparisons with in situ observations from WIND, MAVEN, and Tianwen-1, as shown in Figure~\ref{fig:validation}. 
Overall, the simulated solar wind reproduces the arrival time, duration, and major plasma signatures of the CIR at all spacecraft locations. 
The characteristic increase in solar wind speed, the compression of plasma density, and the enhancement of magnetic field strength are consistently reproduced. 
Although moderate discrepancies remain in magnetic field components, the overall agreement demonstrates that the present MHD model successfully reconstructs the large-scale heliospheric environment and accurately captures the propagation of the target CIR. The validated simulation results therefore provide a reliable basis for investigating the global three-dimensional morphology and radial evolution of the target CIR.

The Wind, MAVEN and Tianwen-1 measurements show a broad compression bounded by gradual stream interfaces rather than the abrupt and correlated jumps in magnetic field strength, plasma density, temperature and flow velocity. No clear shock structures were identified in the in situ measurements at either 1 AU or about 1.6 AU, as also shown in \citet{Zhang2024}. The simulated interaction region is consistent with these observations: the compression develops pronounced gradients, but no well-defined shock forms within the simulated radial range. 

\subsection{Geometry of the CIR}

\begin{figure*}[ht!]
\centering
\includegraphics[width=\textwidth]{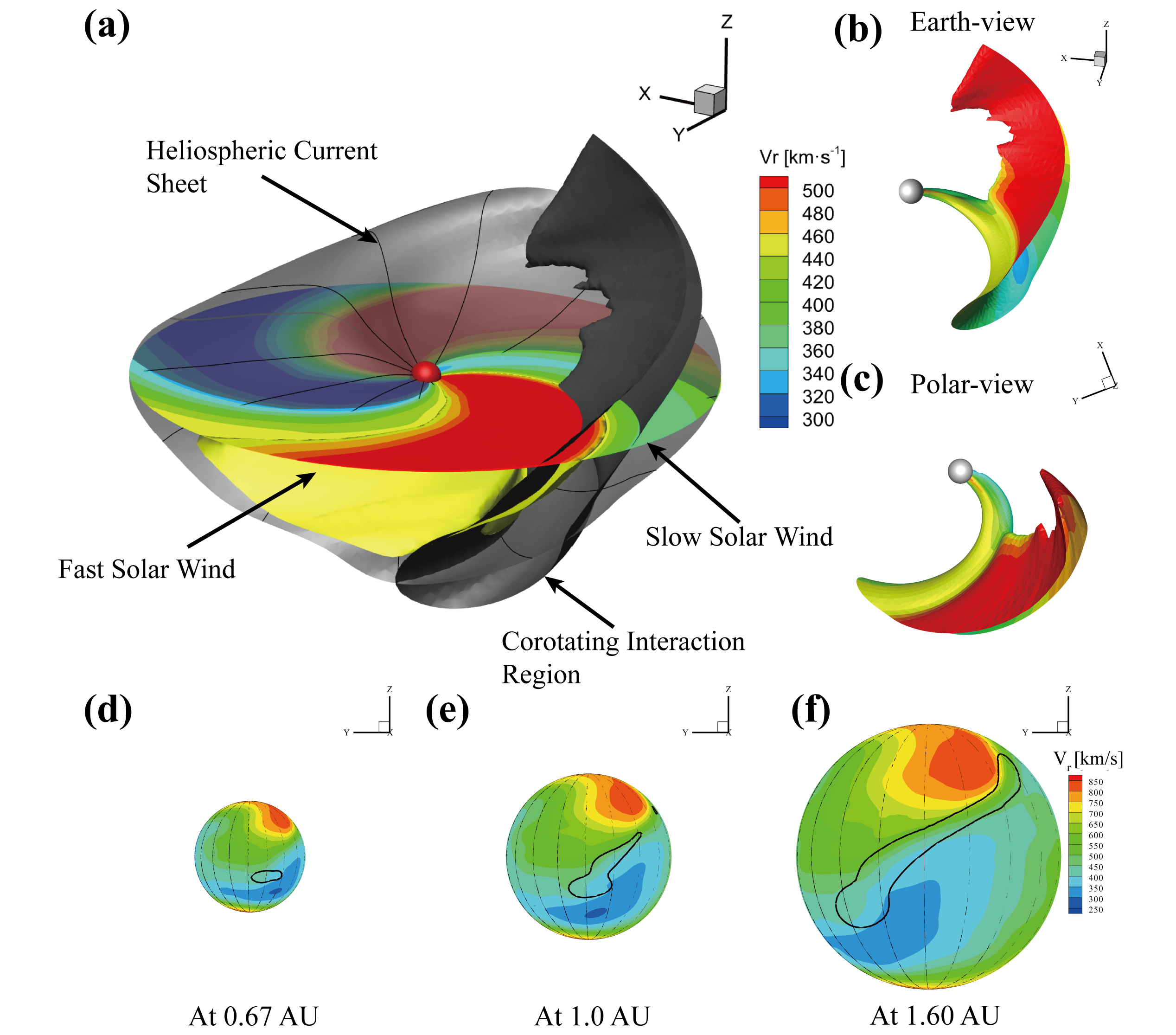}
\caption{3D morphology of the target CIR. Panel (a) shows the overall simulation domain, where the CIR is rendered as a dark gray surface representing the iso-surface of the enhanced total pressure. The CIR is embedded in the background solar wind, which is color-coded by radial velocity; the translucent surface denotes the heliospheric current sheet (HCS), and black lines indicate selected streamlines. Panels (b)-(c) present the Earth-view and polar-view perspectives of the CIR, respectively, with the CIR identified by the iso-surface of the enhanced total pressure and color-coded by radial velocity. Panels (d)-(f) display the radial velocity distributions at heliocentric distances of 0.67~AU, 1.0~AU, and 1.60~AU, corresponding to the locations of BepiColombo, Wind, and MAVEN/Tianwen-1, respectively. The black curves in panels (d)-(f) mark the cross-section of the target CIR.}
\label{fig:3dstructure}
\end{figure*}

Figure~\ref{fig:3dstructure} illustrates the 3D morphology of the target CIR, identified by the enhanced total pressure. Panel (a) presents the overall simulation domain, where the CIR is depicted as a dark gray isosurface of the total pressure  embedded in the background solar wind. The CIR forms a shell-like structure extending along the interface between the fast and slow solar wind streams. It is tilted relative to the HCS, with its lower portion attached to the HCS and the remaining part gradually extending toward higher latitudes. This topology is consistent with the conceptual picture proposed by previous theoretical studies \citep{Riley2007}.

Panels (b)-(c) provide Earth-view and polar-view perspectives of the CIR, respectively, offering a more complete visualization of its global morphology throughout the inner heliosphere. They reveal a warped, shell-like spiral structure that extends over a broad longitudinal and latitudinal range. The color coding by radial velocity highlights the velocity gradient across the CIR, with the highest speeds on the trailing side and lower speeds on the leading flank, consistent with typical CIR flow patterns.

Panels (d)-(f) display the radial velocity distributions at heliocentric distances of 0.67~AU, 1.0~AU, and 1.60~AU, with the black curves marking the cross-section of the CIR at each distance. As the CIR propagates outward, its morphology undergoes clear radial evolution: at 0.67~AU, it appears as a relatively narrow and modest compression region; at 1.0~AU, it becomes more prominent with a broader extent and sharper gradients; and at 1.60~ AU, it has expanded significantly in both latitudinal and longitudinal dimensions, reflecting its development into a fully evolved interaction region with well-defined forward and reverse boundaries.  

\begin{figure*}[ht!]
\centering
\includegraphics[width=\textwidth]{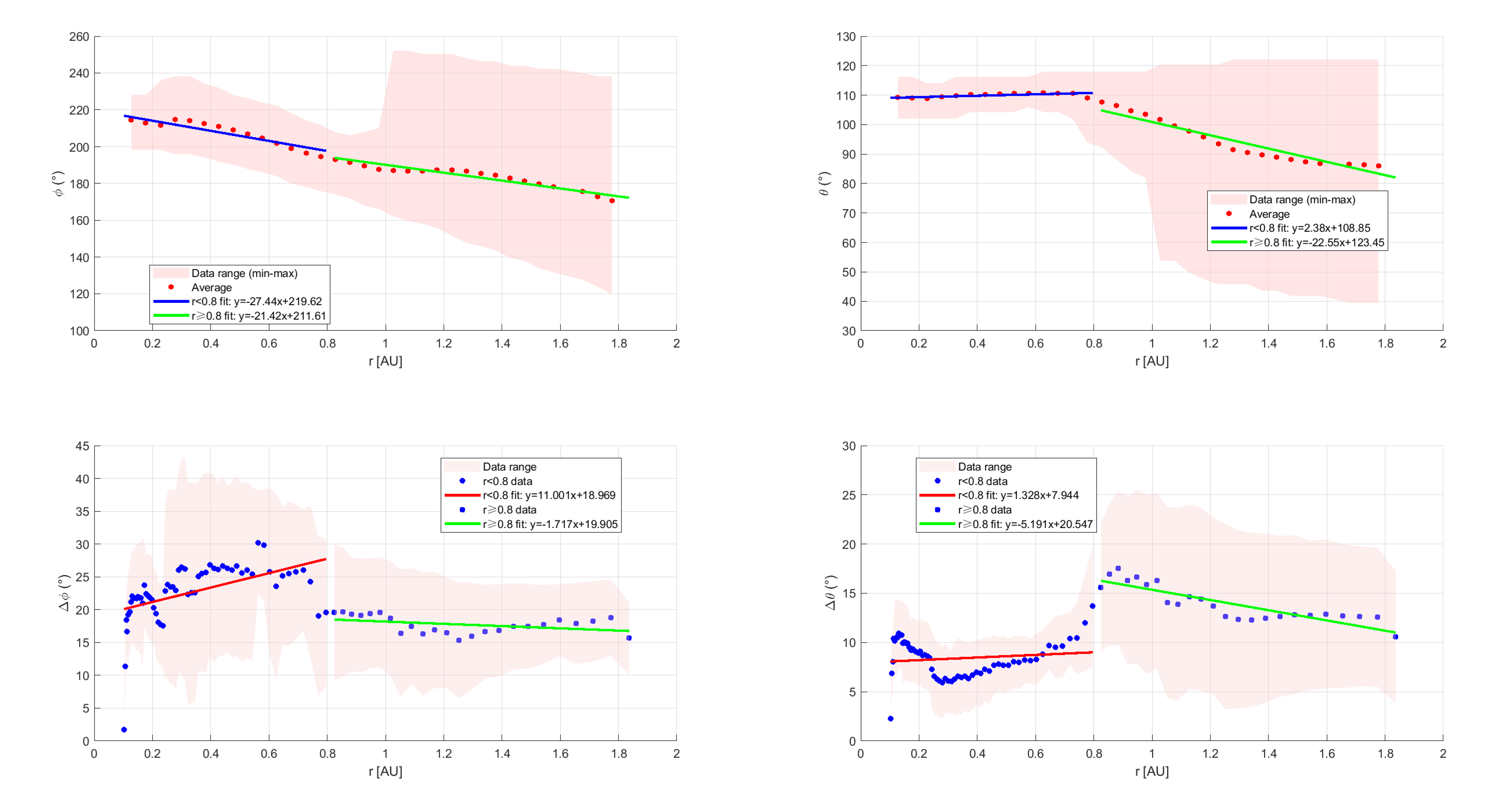}
\caption{Radial variation of the CIR longitude ($\phi$), co-latitude ($\theta$), the longitude width $\Delta \phi$  on a spherical surface of constant radius at a given latitude, and the latitude width $\Delta \theta$ on a spherical surface of constant radius at a given  longitude. Columns labeled ``data" list the values directly extracted from the simulation, with the red and blue points indicating the averaged values at each radial distance.  Linear fits are applied to the averaged data.}
\label{fig:geometry}
\end{figure*} 

The geometric parameters of the CIR are further quantified by tracking their spatial extent at different heliocentric distances. Figure~\ref{fig:geometry} presents the radial variation of the CIR geometry, showing the longitude $\phi$ and colatitude $\theta$ of the CIR as well as their corresponding widths $\Delta\phi$ and $\Delta\theta$ as functions of heliocentric distance $r$.  Within approximately 0.8~AU, the interaction between the trailing high-speed stream and the leading slow solar wind occur gradually, which causes the CIR to shift progressively in the longitudinal direction, while its latitude remains nearly unchanged. Meanwhile, the widths $\Delta\phi$ and $\Delta\theta$ show only a slight increase. Beyond approximately 0.8~AU, the interaction is deemed to intensify significantly. The CIR continues to drift in the longitudinal direction, with its longitudinal extent increasing steadily. As $r$ increases, the colatitude of the CIR decreases overall, indicating that the CIR expands from low latitudes toward higher latitudes as it propagates outward. At the same time, the widths $\Delta\phi$ and $\Delta\theta$ exhibit a slight decrease with radial distance, implying that the CIR becomes increasingly narrow in its angular extent and progressively more compressed as it propagates outward. This behavior is consistent with the scenario in which the continuous interaction between the fast and slow solar wind streams leads to accumulating compression effects, channeling the enhanced pressure and density into a progressively thinner shell-like structure.

\subsection{Radial Evolution of the CIR Parameters}

\begin{figure*}[ht!]
\centering
\includegraphics[width=\textwidth]{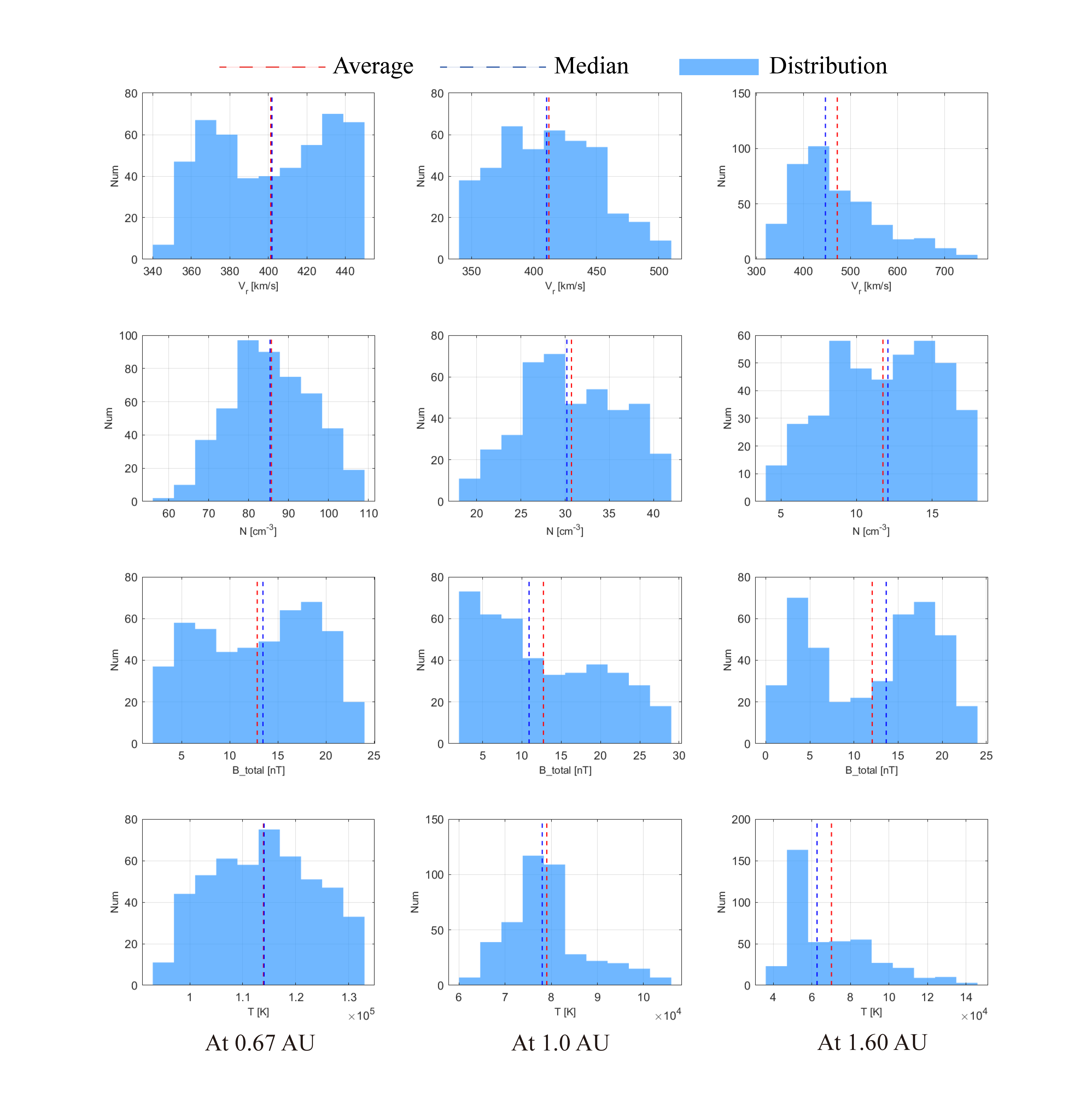}
\caption{Statistical distributions of radial velocity $V_r$ (in km~s$^{-1}$), number density $N$ (in cm$^{-3}$), total magnetic field strength $B_\textrm{total}$ (in nT), and temperature $T$ (in K) inside the CIR at heliocentric distances of 0.67~AU, 1.0~AU, and 1.60~AU. The red and blue dashed lines denote the mean and median values of each distribution, respectively.}
\label{fig:parameter_profile}
\end{figure*}

Figure~\ref{fig:parameter_profile} presents the statistical distributions of radial velocity $V_r$, number density $N$, total magnetic field strength $B_\textrm{total}$, and temperature $T$ inside the CIR at heliocentric distances of 0.67~AU, 1.0~AU, and 1.60~AU. The distributions are computed from the values at the individual grid points of the CIR: each grid point inside the CIR contributes one sample, and at every radial distance all grid points of the CIR, over its full longitudinal and latitudinal extent, are included, so that the statistics is a sampling of the entire CIR at every radial distance. The volume element over which the CIR is divided is approximately 2.2, 3.4 and 5.8~Rs$^3$ at 0.67, 1.0 and 1.6~AU, respectively.

The distributions reveal distinct evolutionary trends among different plasma parameters as the CIR propagates outward. For the radial velocity, the distribution transitions from a double-peaked profile at 0.67~AU to a single-peaked profile at 1.60~AU, indicating that the high-speed solar wind is decelerated while the low-speed solar wind is accelerated during CIR propagation, eventually leading to a more monotonically distributed velocity within the interaction region. However, the distribution of the total magnetic field strength evolves into a clear double-peaked structure at larger distances, suggesting that the initially weak magnetic field within the CIR becomes increasingly compressed as the interaction intensifies, resulting in a growing fraction of the CIR volume occupied by strengthened magnetic field. For the plasma properties, however, the compression effects are less pronounced. The distributions of number density and temperature show only modest changes with radial distance, and do not develop the double-peaked feature seen in the magnetic field. This contrast implies that the magnetic field is more sensitive to the interaction process than the plasma density and temperature, likely due to the different response of these quantities to the dynamic pressure balance and the geometry of the compression region as the CIR evolves outward.

\begin{figure*}[ht!]
\centering
\includegraphics[width=\textwidth]{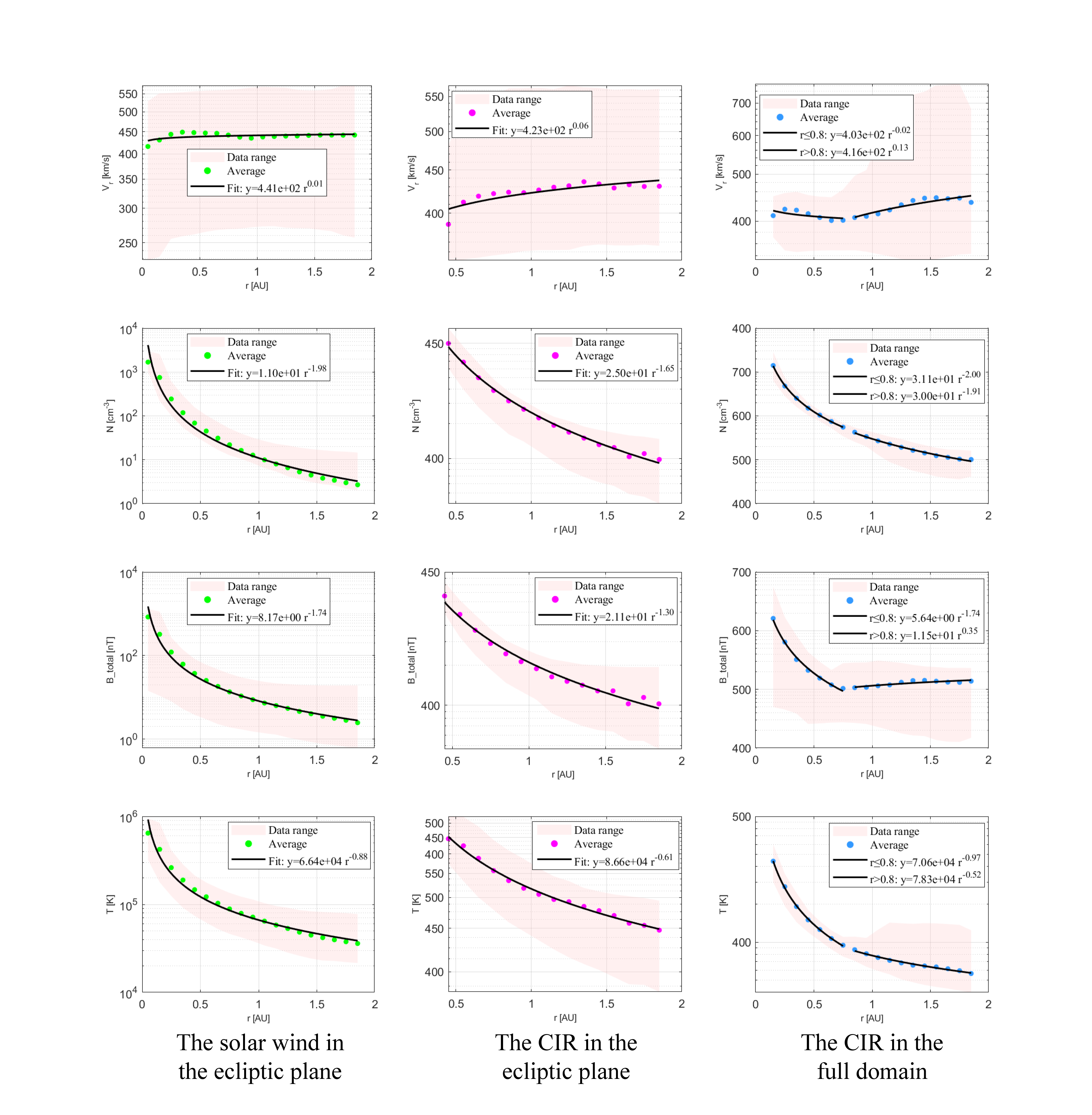} 
\caption{Radial evolution of solar wind and CIR plasma parameters, including radial velocity $V_r$ (in km~s$^{-1}$), number density $N$ (in cm$^{-3}$), total magnetic field strength $B_\textrm{total}$ (in nT), and temperature $T$ (in K). The left, middle, and right panels show the results for the solar wind in the ecliptic plane, the CIR in the ecliptic plane, and the CIR in the full domain, respectively. The scattered points represent values directly extracted from the simulation at each radial distance, with the green, pink, and blue symbols denoting the averaged values. The solid black curves correspond to power-law fits to the averaged data.}
\label{fig:plasma_fits} 
\end{figure*}

Figure~\ref{fig:plasma_fits} presents the radial evolution of plasma parameters for three different contexts: the background solar wind in the ecliptic plane (left), the CIR in the ecliptic plane (middle), and the CIR in the full 3D domain (right). For the background solar wind in the ecliptic plane, the power-law fits yield radial scalings of the radial velocity $V_r \propto r^{0.01}$, of the number density $N \propto r^{-1.98}$, of the total magnetic field strength $B_\textrm{total} \propto r^{-1.74}$, and of the temperature $T \propto r^{-0.88}$, indicating that the radial velocity remains nearly constant with distance, while the density and magnetic field decrease monotonically as expected for an expanding solar wind. These scaling relations are consistent with previous observational and modeling studies of the ambient solar wind \citep[e.g.,][]{Venzmer2018, Geyer2021, Scolini2021EUHFORIA}.

For the CIR in the ecliptic plane, the fitted exponents are $V_r \propto r^{0.06}$, $N \propto r^{-1.65}$, $B_\textrm{total} \propto r^{-1.30}$, and $T \propto r^{-0.61}$. Compared with the background solar wind, the CIR in the ecliptic plane shows a slightly less steep decrease in density, magnetic field strength, and temperature, albeit with moderate modifications due to compression effects. Overall, however, the differences between the CIR and the background solar wind scalings are not substantial in the ecliptic plane.

For the CIR in the full domain, the radial evolution exhibits two distinct regimes separated at approximately 0.8~AU, similar to the geometric parameters of the CIR shown in Figure~\ref{fig:geometry}. Within about 0.8~AU, the scaling exponents are $V_r \propto r^{0.02}$, $N \propto r^{-2.00}$, $B_\textrm{total} \propto r^{-1.74}$, and $T \propto r^{-0.97}$. These exponents are comparable to those of the background solar wind in the ecliptic plane, indicating that the early-stage interaction has little influence on the radial evolution of the plasma and magnetic field. Beyond approximately 0.8~AU, where the interaction is deemed to intensify significantly, the scaling exponents become $V_r \propto r^{0.13}$, $N \propto r^{-1.91}$, $B_\textrm{total} \propto r^{0.35}$, and $T \propto r^{-0.52}$. This second regime shows a marked acceleration in the radial decline of all parameters, particularly for the magnetic field and temperature, reflecting the continuous conversion of the kinetic energy of the fast solar wind into the magnetic and thermal energies at the stream interface. Consequently, the CIR maintains enhanced density, magnetic field strength, and temperature over a large radial distance. These full-domain scalings differ notably from those obtained in the ecliptic plane, suggesting that the inclusion of the full latitudinal and longitudinal extent of the CIR captures additional compression effects that are not fully represented in the ecliptic-plane cut alone.

\section{Summary and Discussion}

In this study, we have investigated the radial evolution of a CIR during CR2251 using the 3D SIP-AMR-CESE MHD model, combined with a two-step backmapping method. By identifying both the origins and the propagation paths of the high-speed solar wind measured by WIND, we introduce a way to reproduce the high-speed solar wind in the MHD model. The AMR technique enables us to capture the CIR with high resolution while maintaining computational efficiency. The simulation successfully reproduces the features of the CIR observed by Wind at 1 AU, and MAVEN and Tianwen-1 at about 1.6 AU. The validated simulation therefore provides a comprehensive basis for analyzing both the geometric and plasma properties of a single CIR throughout its evolution from the inner heliosphere to beyond 1.6 AU. Our main findings are summarized as follows.

Regarding the geometric properties of the CIR, its three-dimensional morphology reveals a warped, shell-like spiral structure tilted relative to the HCS, with its lower portion attached to the HCS and the remaining part gradually extending toward higher latitudes. At 0.67~AU, the CIR appears as a relatively narrow and modest compression region; at 1.0~AU, it becomes more prominent with a broader extent and sharper gradients; and at 1.60~AU, it has expanded significantly in both longitudinal and latitudinal dimensions. This topology is consistent with the conceptual picture proposed by \citet{Riley2007}. As the CIR propagates outward, it drifts progressively in longitude, with its longitudinal extent increasing steadily, while simultaneously expanding from low latitudes toward higher latitudes. Beyond approximately 0.8~AU, both the longitudinal and latitudinal widths exhibit a slight decrease with radial distance, implying that the CIR becomes increasingly narrow and more compressed in its angular extent.

For the radial evolution of the CIR parameters, the radial velocity distribution transitions from a double-peaked profile at 0.67~AU to a single-peaked profile at 1.60~AU, reflecting the deceleration of the high-speed solar wind and the acceleration of the low-speed solar wind as they interact within the CIR. In contrast, the total magnetic field strength evolves into a clear double-peaked structure at larger distances, indicating that an increasing fraction of the CIR volume is occupied by compressed and strengthened magnetic field. The radial evolution of plasma parameters is found to differ between the ecliptic-plane and full-domain analyses. In the ecliptic plane, the radial scalings for the background solar wind ($V_r \propto r^{0.01}$, $N \propto r^{-1.98}$, $B_{\rm total} \propto r^{-1.74}$, $T \propto r^{-0.88}$) are consistent with previous observational and modeling studies. The CIR in the ecliptic plane exhibits power-law scalings similar to those of the background solar wind. However, the full-domain analysis reveals a distinct two-regime behavior: within about 0.8~AU, the scalings ($V_r \propto r^{0.02}$, $N \propto r^{-2.00}$, $B_{\rm total} \propto r^{-1.74}$, $T \propto r^{-0.97}$) are comparable to those of the background solar wind; beyond about 0.8~AU, the scalings change markedly to $V_r \propto r^{0.13}$, $N \propto r^{-1.91}$, $B_{\rm total} \propto r^{0.35}$, and $T \propto r^{-0.52}$.  

The two-stage evolutionary picture identified in this study provides a quantitative framework for understanding how CIRs develop from nascent compression regions into fully evolved interaction structures. The transition at about 0.8~AU, evident in both geometric parameters and plasma scalings, suggests a critical distance where the cumulative effect of stream interaction becomes dynamically significant. This result is consistent with the observational finding that the occurrence rate of CIRs with sharp stream interfaces increases substantially between 0.7 and 1~AU \citep{Jian2008}, and with the higher fast forward shock occurrence rate observed at Mars compared to Earth \citep{Henderson2025}.

The significant differences between ecliptic-plane and full-domain scalings underscore the limitations of relying solely on single-point or ecliptic-only measurements. While ecliptic-plane observations have provided valuable insights into CIR properties, they may underestimate the full extent of compression and the three-dimensional nature of the interaction region. The full-domain analysis suggests that the latitudinal extent of CIRs plays a crucial role in their evolution, and that the compression effects are more pronounced when the entire three-dimensional structure is considered. This finding is particularly relevant for space weather forecasting (away from Earth), as it implies that predictions based on one-dimensional or ecliptic-only models may not fully capture the intensity of CIRs at different heliographic latitudes.

Several aspects of this study entail further investigation. The present simulation focuses on a single CIR event; whether the two-stage evolution, its transition position at about 0.8~AU, and the specific scaling exponents are universal or event-dependent remains to be tested through simulations of additional events under different solar wind conditions. What is the role of the HCS geometry in determining the three-dimensional topology of CIRs and how this topology influences the radial evolution of plasma parameters, deserves systematic exploration. Furthermore, the CIR is identified according to the total-pressure enhancement derived from the MHD simulation. Although this criterion effectively captures the large-scale compression region, future work is needed to investigate whether different identification methods could lead to differences in the inferred boundaries of the CIR.

From a methodological perspective, the combination of source mapping, extra solar wind heating, and AMR-based MHD simulation has proven effective in capturing the detailed evolution of a single CIR. This approach could be extended to the study of other heliospheric structures, such as CME-CIR interactions and merged interaction regions, which are known to play important roles in the transport of energetic particles and the modulation of galactic cosmic rays in the inner and outer heliosphere.

In summary, this study provides a comprehensive, three-dimensional characterization of the radial evolution of a CIR from its formation to its maturity beyond 1.6~AU. The quantitative scalings and two-stage evolutionary framework established here offer a useful reference for interpreting past observations, guiding future observational campaigns, and improving space weather prediction models. The identified differences between the ecliptic-plane and full-domain scalings highlight the importance of considering the full three-dimensional structure of CIRs in both observational and modeling studies. Future out-of-ecliptic measurements from Solar Orbiter \citep{Muller2020}, China's planned Solar
Polar-orbit Observatory \citep{Deng2025}, and solar magnetic field measurements \citep{2020Yangb, 2024Yangc, 2026DengY} are expected to provide direct observational constraints on the three-dimensional morphology and latitudinal structure of CIRs, complementing the simulation-based findings of this study.
 
\section*{acknowledgments}
  
This work is supported by the National Key R\&D Program of China (No. 2022YFF0503800), the National Natural Science Foundation of China (NSFC) (Nos. 42330210, 42474216, 42274213), China's Space Origins Exploration Program, and Project Supported by the Specialized Research Fund for State Key Laboratory of Solar Activity and Space Weather. The work was carried out at National Supercomputer Center in Tianjin, China, and the calculations were performed on TianHe-1 (A). We would like to thank the use of data from the SOHO and STEREO spacecraft. Tianwen-1 magnetic field data are available through the Planet Exploration Program Scientific Data Release System (\url{http://202.106.152.98:8081/marsdata/}) or the official website of the MOMAG team (\url{http://space.ustc.edu.cn/dreams/tw1_momag/}). MAVEN data are available from NASA's Planetary Data System (\url{https://pds-ppi.igpp.ucla.edu/mission/MAVEN/MAVEN/}  \citet{https://doi.org/10.17189/6jmv-t325}).
.



\bibliography{sample701}{}
\bibliographystyle{aasjournalv7}



\end{document}